# Activation cross sections of $\alpha$-induced reactions on $^{nat}Zn$ for Ge and Ga production

M. Aikawa[1], M. Saito[2*], S. Ebata[1†], Y. Komori[3], H. Haba[3]

[1] Faculty of Science, Hokkaido University, Sapporo 060-0810, Japan

[2] Graduate School of Science, Hokkaido University, Sapporo 060-0810, Japan

[3] Nishina Center for Accelerator-Based Science, RIKEN, Wako 351-0198, Japan

The production cross sections of $^{68,69}$Ge and $^{66,67}$Ga by $\alpha$-induced reactions on $^{nat}$Zn have been measured using the stacked-foil activation method and off-line $\gamma$-ray spectrometry from their threshold energies to 50.7 MeV. The derived cross sections were compared with the previous experimental data and the calculated values in the TENLD-2017 library. Our result shows a slightly larger amplitude than the previous data at the peak, though the peak energy is consistent with them.

## 1. Introduction

A variety of radioactive isotopes are used in nuclear medicine for therapy and diagnosis. One of such isotopes, $^{68}$Ga ($T_{1/2}$ = 67.71 min), is available for the positron emission tomography (PET) [1]. The production of $^{68}$Ga is an important topic for the application. For the practical application, however, the half-life of about one hour is too short to deliver $^{68}$Ga from production facilities to hospitals outside. In addition to $^{68}$Ga, therefore, the production of its long-lived parent, $^{68}$Ge ($T_{1/2}$ = 270.95 d), is worthy of investigation for a $^{68}$Ga generator [2].

There are several possible reactions to produce $^{68}$Ge. One of the reactions is $\alpha$-induced reactions on Zn isotopes. The capture reactions of $\alpha$-particles on $^{64}$Zn have energetically been investigated due to interests in astrophysics [3,4] and nuclear physics [5–8]. The cross sections of the capture reaction are less than a few mb. On the other hand, the cross sections to produce $^{68}$Ge by the $\alpha$-induced reactions on $^{nat}$Zn reach approximately 150 mb at the peak around the incident energy of 30 MeV [9,10]. However, the energy-dependent shapes of the cross sections are slightly different from each other. We therefore measured the cross sections of the $\alpha$-induced reactions on $^{nat}$Zn for the $^{68}$Ge production. From the measured data, the integral

[*] Present address: Graduate School of Biomedical Science and Engineering, Hokkaido University, Sapporo 060-8638, Japan

[†] Present address: School of Environment and Society, Tokyo Institute of Technology, Tokyo 152-8550, Japan

yield of $^{68}$Ge was derived. The cross sections for the long-lived nuclides, $^{69}$Ge and $^{66,67}$Ga, were also measured and compared with the earlier experimental data [9,10].

## 2. Experimental details and data analysis

The experiment was performed at the AVF cyclotron of the RIKEN RI Beam Factory by using the stacked foil activation method and off-line γ-ray spectrometric technique. Thin metallic foils of $^{nat}$Zn (purity: 99.9%, Nilaco Corp., Japan) and $^{nat}$Ti (purity: 99.6%, Nilaco Corp., Japan) were used. The thicknesses of the Zn and Ti foils were estimated from measured area and weight of the foils (approximately 50 × 50 mm$^2$ and 50 × 100 mm$^2$) and found to be 18.64 mg/cm$^2$ and 2.25 mg/cm$^2$, respectively. The stacked target consisted of 14 sets of the Zn-Ti-Ti foils (8 × 8 mm$^2$) cut off from the large foils. The Zn and the first Ti foils to catch recoil products were measured together to derive the cross sections of products formed in the Zn foils. Losses of products in the second Ti foils at the downstream of the beam were compensated from the first Ti foils and only the second foils were measured for the $^{nat}$Ti(α,x)$^{51}$Cr monitor reaction. The stacked target was irradiated for 2 hours by a 51.5 MeV alpha beam with the average intensity of 41.0 pnA, which was measured by a Faraday cup like target holder. The incident beam energy was measured by the time-of-flight method using a plastic scintillator monitor [11]. The energy of projectiles decreasing in the foils was estimated using the SRIM code [12]. The γ-lines from the irradiated foils were measured by HPGe detectors (ORTEC GEM-25185-P and ORTEC GEM35P4-70) and analyzed by Gamma Studio (SEIKO EG&G) for the high resolution γ-ray spectrometry. The detectors were calibrated by a multiple standard gamma-ray point source, which consisted of, $^{57,60}$Co, $^{88}$Y, $^{109}$Cd, $^{113}$Sn $^{137}$Cs, $^{139}$Ce and $^{241}$Am. The reaction and decay data for the γ-ray spectrometry were taken from Nudat 2.7 [13] and QCalc [14] and summarized in Table 1.

The activation cross sections σ were deduced using the standard activation formula

$$\sigma = \frac{C_\gamma \lambda}{\varepsilon_d \varepsilon_\gamma \varepsilon_t N_t N_b (1 - e^{-\lambda t_b}) e^{-\lambda t_c} (1 - e^{-\lambda t_m})} \tag{1}$$

where $C_\gamma$ is the measured net counts of the peak area, $\lambda$ is the decay constant (s$^{-1}$), $\varepsilon_d$ is the detector efficiency, $\varepsilon_\gamma$ is the gamma-ray abundance, $\varepsilon_t$ is the measurement dead time, $N_t$ is the surface density of target atoms (cm$^{-2}$), $N_b$ is the number of bombarding particles per unit time (s$^{-1}$), $t_b$ is the bombarding time (s), $t_c$ is the cooling time (s), and $t_m$ is the acquisition time (s).

The integral yield $Y(E_{\text{in}})$ (kBq μA$^{-1}$ h$^{-1}$) was estimated using the equation

$$Y(E_{\text{in}}) = I \frac{(1 - e^{-\lambda T})}{\lambda} \int_0^{E_{\text{in}}} \frac{\sigma(E)}{S(E)} dE \tag{2}$$

where $E_{\text{in}}$ is the incident energy (MeV), $I$ is the unit beam intensity (1 μA) (s$^{-1}$), $T$ is the unit beam irradiation period (1 h) (s), $\sigma(E)$ is the cross sections (cm$^2$), $S(E)$ is the stopping power (MeV cm$^2$).

To assess beam parameters, the cross sections of the $^{nat}$Ti(α,x)$^{51}$Cr monitor reaction were derived by measuring the 320.08-keV γ-ray ($I_\gamma$ = 9.91%) from the $^{51}$Cr decay ($T_{1/2}$ = 27.7025 d) after a cooling time of 21 hours for background reduction. The result is in good agreement with the recommended values [15] as shown in Fig. 1. Therefore, we regard that any adjustments of the measured beam parameters were not necessary.

Table 1: Reaction and decay data taken from online databases [13,14]

| Nuclide | Half-life | Decay mode (%) | $E_\gamma$ (keV) | $I_\gamma$ (%) | Contributing reaction | Q-value (MeV) |
|---|---|---|---|---|---|---|
| $^{69}Ge$ | 39.05 h | $\varepsilon+\beta^+$ (100) | 574.11 | 13.3(18) | $^{66}Zn(\alpha,n)$ | -7.4 |
| | | | 871.98 | 11.9(16) | $^{67}Zn(\alpha,2n)$ | -14.5 |
| | | | 1106.77 | 36 | $^{68}Zn(\alpha,3n)$ | -24.7 |
| | | | | | $^{70}Zn(\alpha,5n)$ | -40.4 |
| $^{68}Ge$ | 270.95 d | $\varepsilon$ (100) | - | - | $^{64}Zn(\alpha,\gamma)$ | 3.4 |
| | | | | | $^{66}Zn(\alpha,2n)$ | -15.6 |
| | | | | | $^{67}Zn(\alpha,3n)$ | -22.7 |
| | | | | | $^{68}Zn(\alpha,4n)$ | -32.9 |
| | | | | | $^{70}Zn(\alpha,6n)$ | -48.6 |
| $^{68}Ga$ | 67.71 min | $\varepsilon+\beta^+$ (100) | 1077.34 | 3.22 | $^{68}Ge(\varepsilon)$ | |
| $^{67}Ga$ | 3.2617 d | $\varepsilon$ (100) | 93.310 | 38.81(3) | $^{64}Zn(\alpha,p)$ | -4.0 |
| | | | 184.576 | 21.410(10) | $^{66}Zn(\alpha,t)$ | -14.5 |
| | | | 300.217 | 16.64(12) | $^{67}Zn(\alpha,tn)$ | -21.6 |
| | | | | | $^{68}Zn(\alpha,t2n)$ | -31.8 |
| | | | | | $^{70}Zn(\alpha,t4n)$ | -47.5 |
| | | | | | $^{67}Ge(\varepsilon)$ | |
| $^{66}Ga$ | 9.49 h | $\varepsilon+\beta^+$ (100) | 1039.220 | 37.0(20) | $^{64}Zn(\alpha,d)$ | -13.0 |
| | | | | | $^{66}Zn(\alpha,tn)$ | -25.8 |
| | | | | | $^{67}Zn(\alpha,t2n)$ | -32.8 |
| | | | | | $^{68}Zn(\alpha,t3n)$ | -43.0 |
| | | | | | $^{66}Ge(\varepsilon)$ | |

## 3. Results and Discussion

The production cross sections for the long-lived Ge and Ga isotopes, $^{68,69}Ge$ and $^{66,67}Ga$, were presented in Table 2 and shown in Figs. 2-5 in comparison with experimental data studied earlier [9,10] and the TENDL-2017 data [16]. In addition, the integral yield of $^{68}Ge$ was derived from the measured cross sections and shown in Fig. 6.

The total uncertainty of the measured cross sections was estimated to be less than 28.3% including statistical errors (<27.5%). It was estimated as the square root of the quadratic summation of the components; the beam intensity (5%), target thickness (1%), target purity (1%), detector efficiency (5%), γ intensity (<5.4%) and peak fitting (3%).

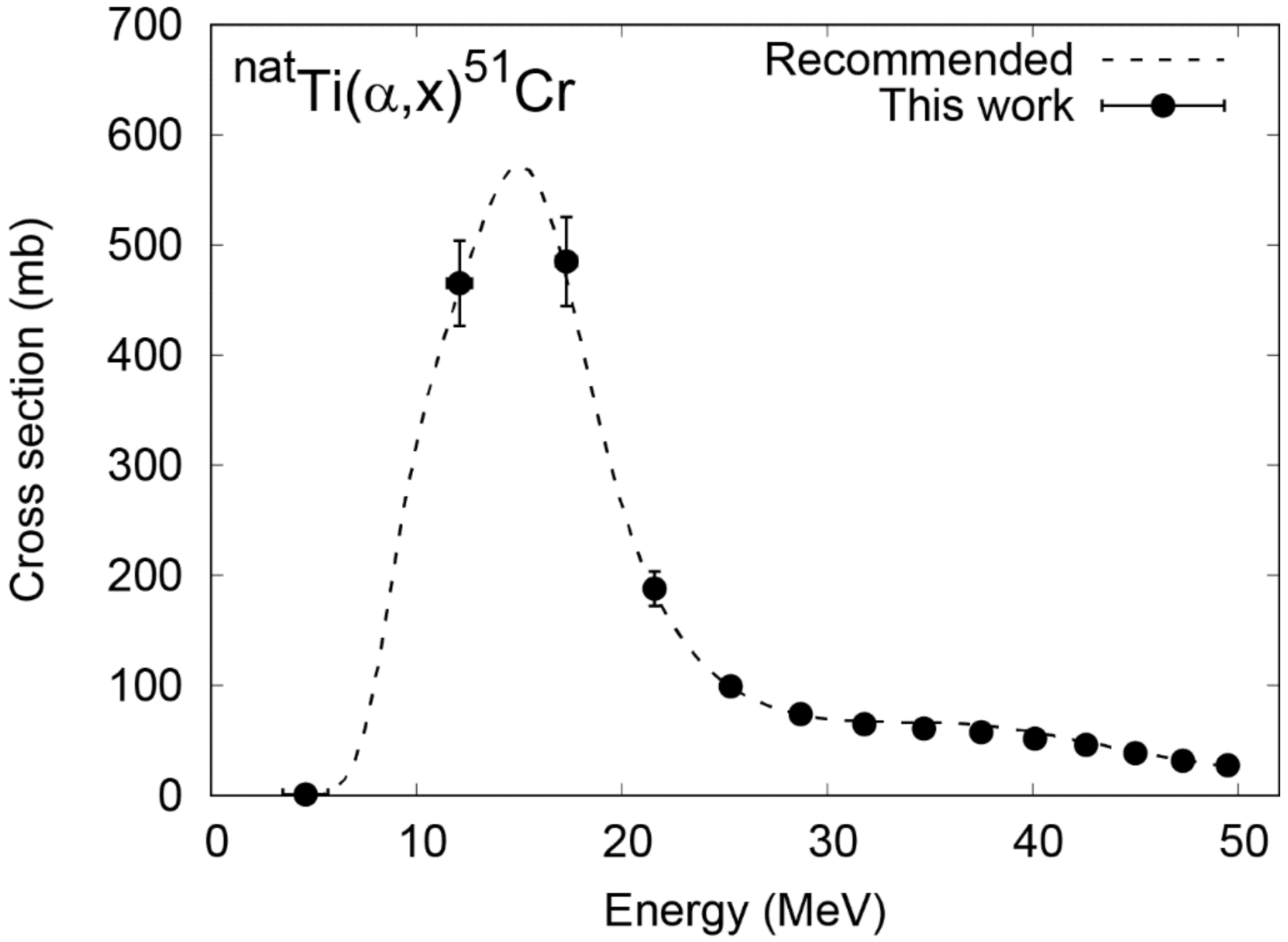

Fig. 1. Measured excitation function of the monitor $^{nat}Ti(\alpha,x)^{51}Cr$ reaction in comparison with the recommended values [15].

Table 2. Measured cross sections for the long-lived Ge and Ga isotope productions.

| Energy (MeV) | $^{nat}Zn(\alpha,x)^{68}Ge$ (mb) | $^{nat}Zn(\alpha,x)^{69}Ge$ (mb) | $^{nat}Zn(\alpha,x)^{66}Ga$ (mb) | $^{nat}Zn(\alpha,x)^{67}Ga$ (mb) |
|---|---|---|---|---|
| 50.7 ±0.9 | 50.3 ±5.5 | 56.7 ±5.0 | 103.8 ±10.4 | 230.3 ±19.2 |
| 48.5 ±1.0 | 46.6 ±5.3 | 70.2 ±6.2 | 103.6 ±10.3 | 251.9 ±21.0 |
| 46.2 ±1.0 | 52.7 ±5.8 | 80.7 ±7.2 | 111.0 ±11.1 | 254.1 ±21.2 |
| 43.9 ±1.1 | 58.6 ±6.2 | 88.5 ±7.8 | 138.4 ±13.8 | 243.6 ±20.3 |
| 41.5 ±1.1 | 70.8 ±7.2 | 92.8 ±8.2 | 189.3 ±18.8 | 214.0 ±17.9 |
| 38.9 ±1.1 | 95.0 ±9.1 | 88.8 ±7.9 | 261.9 ±26.0 | 167.5 ±14.0 |
| 36.2 ±1.2 | 128.9 ±12.2 | 75.9 ±6.7 | 344.8 ±34.3 | 110.7 ±9.3 |
| 33.4 ±1.3 | 156.2 ±14.5 | 59.4 ±5.3 | 411.4 ±40.9 | 67.3 ±5.7 |
| 30.4 ±1.4 | 176.4 ±16.3 | 46.9 ±4.1 | 408.6 ±40.6 | 64.8 ±5.4 |
| 27.2 ±1.5 | 153.4 ±14.3 | 59.0 ±5.2 | 337.7 ±33.6 | 126.0 ±10.5 |
| 23.6 ±1.6 | 88.5 ±8.5 | 109.4 ±9.7 | 173.2 ±17.3 | 292.1 ±24.4 |
| 19.7 ±1.9 | 23.3 ±2.8 | 162.4 ±14.3 | 17.3 ±2.2 | 457.1 ±38.1 |
| 15.0 ±2.2 | 0.91 ±0.26 | 123.2 ±10.9 | | 337.7 ±28.2 |
| 9.0 ±3.1 | 0.78 ±0.14 | 12.2 ±1.1 | | 28.2 ±2.4 |

### 3.1 $^{68}$Ge production

The measurement of the 1077.34-keV γ-line ($I_γ$ = 3.22%) from the $^{68}$Ga decay ($T_{1/2}$ = 67.71 min), which was in equilibrium with that of its parent $^{68}$Ge ($T_{1/2}$ = 270.95 d), was performed after a long cooling time of about 80 days. In the cooling time, $^{68}$Ga produced directly through the $^{nat}$Zn(α,x) reaction could be considered to decay out completely. The cross sections of $^{68}$Ge are shown in Fig. 2 together with the previous experimental data [9,10] and the TENDL-2017 data [16]. Our result with the spline fit is slightly different from the previous data, though the peak position at around 30 MeV is consistent with them. On the contrary, the TENDL-2017 data shows different tendency from the experimental data. The discrepancy probably comes from underestimation of the contribution from the $^{67}$Zn(α,3n)$^{68}$Ge reaction with the Q-value of -22.7 MeV.

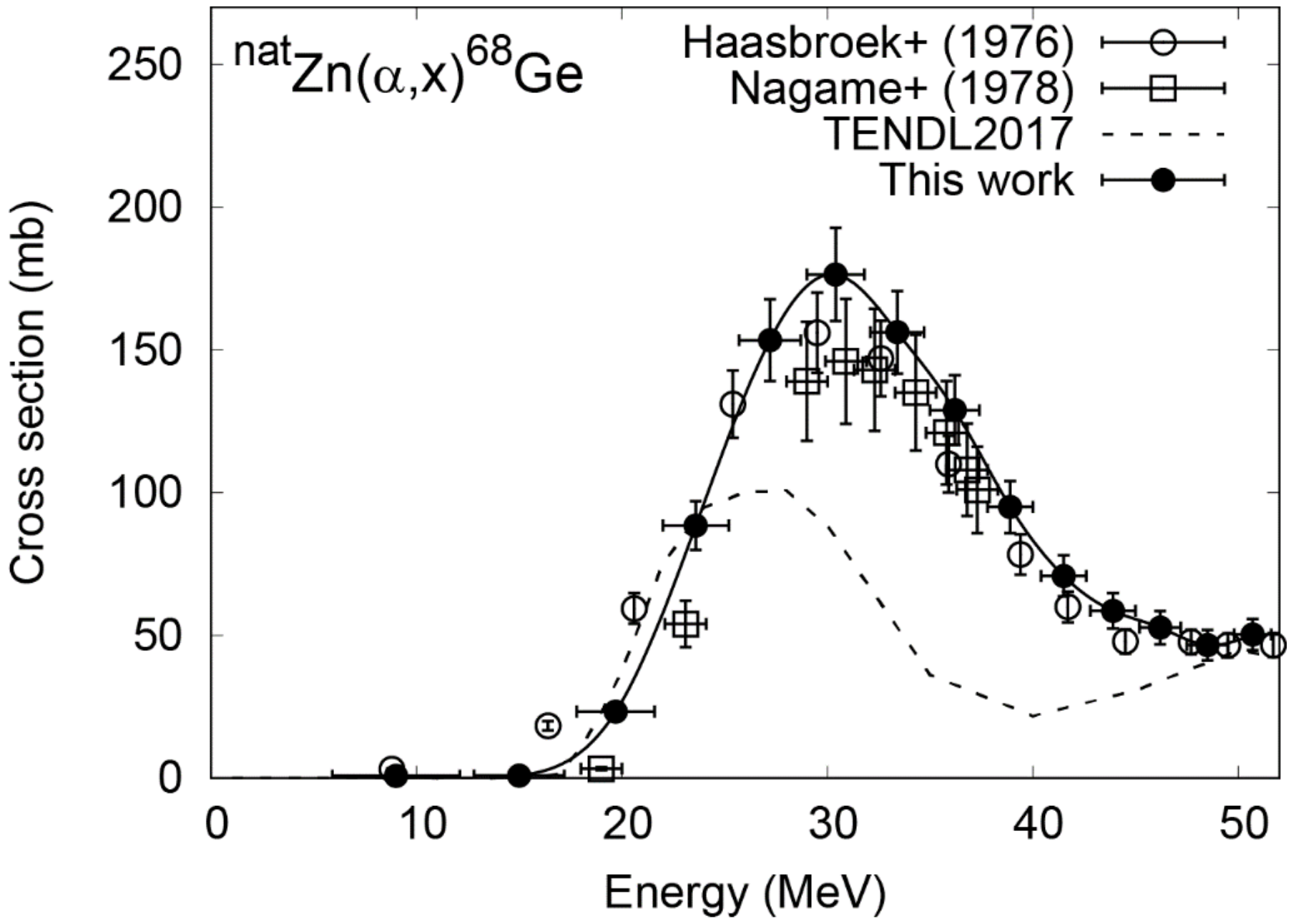

Fig. 2. Measured excitation function of the $^{nat}$Zn(α,x)$^{68}$Ge reaction. The result is compared with the previous experimental data [9,10] and the TENDL-2017 data [16].

### 3.2 $^{69}$Ge production

The cross sections for $^{69}$Ge ($T_{1/2}$ = 39.05 h) production were derived from the measurement of the 1106.77-keV γ-line ($I_\gamma$ = 36%) after the 22-hours cooling time. The contribution of the 1107.3-keV γ-line ($I_\gamma$ = 2.2%) of $^{71m}$Zn ($T_{1/2}$ = 13.76 h) was considered to be negligible, because its more intense γ-line at 386.28 keV ($I_\gamma$ = 93%) was not detected. The excitation function of the $^{nat}$Zn(α,x)$^{69}$Ge reaction is shown in Fig. 3 in comparison with the experimental data studied earlier [10] and the TENDL-2017 data [16]. The present result is in good agreement with the previous data [10], though slightly different from the TENDL-2017 data. The peaks of the TENDL-2017 data shifts to the lower energy side and the cross sections at around 40 MeV are overestimated.

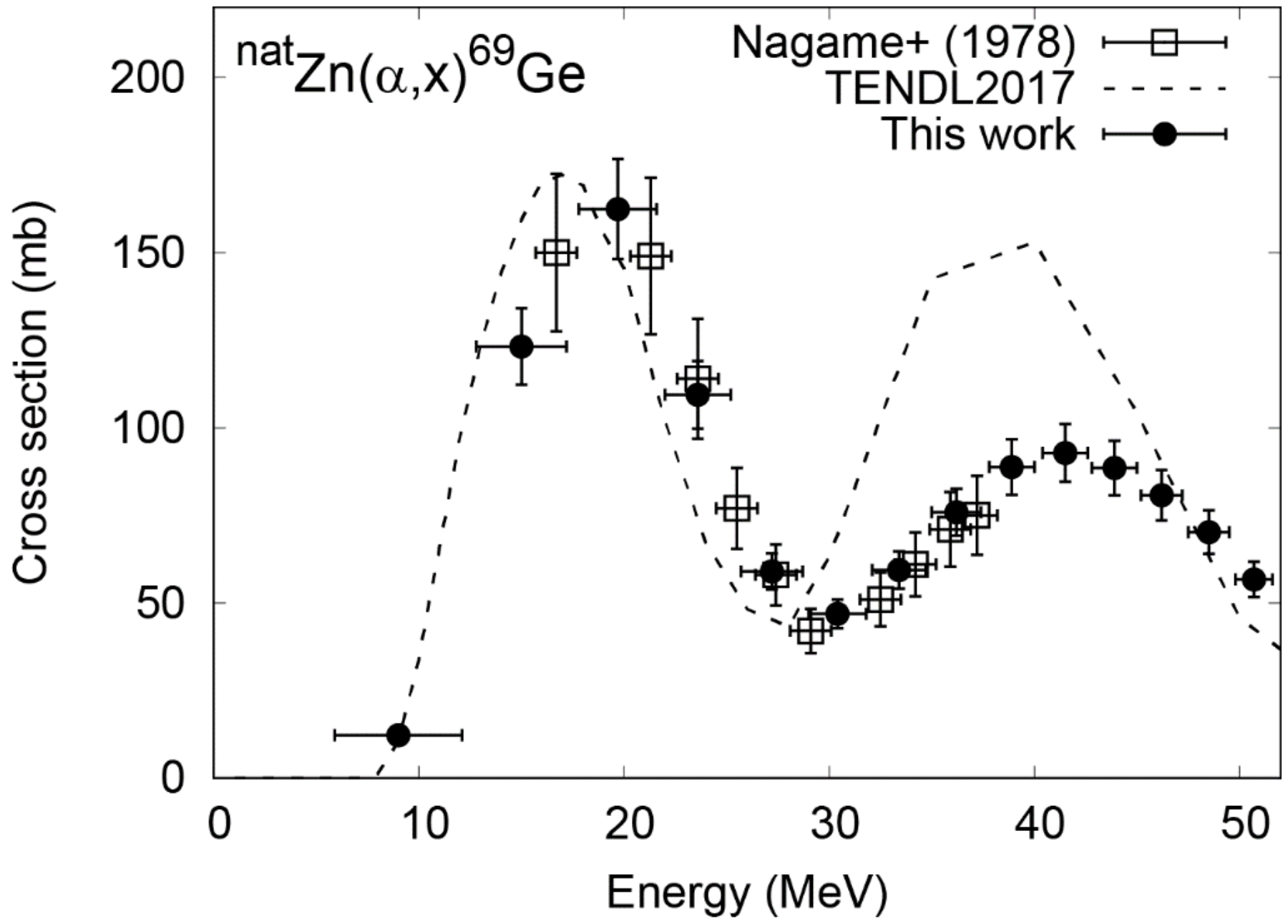

Fig. 3. Measured excitation function of the $^{nat}$Zn(α,x)$^{69}$Ge reaction. The result is compared with the previous experimental data [10] and the TENDL-2017 data [16].

### 3.3 $^{66}$Ga production

The excitation function of the $^{nat}Zn(\alpha,x)^{66}Ga$ reaction was derived from the measurement of the 1039.220-keV γ-line ($I_\gamma$ = 37.0%). The cumulative cross sections for $^{66}$Ga ($T_{1/2}$ = 9.49 h) were derived from the measurement series after a cooling time of 22 hours. The parent nuclide $^{66}$Ge ($T_{1/2}$ = 2.26 h) could be considered to decay in the cooling time. The result is shown in Fig. 4 with the experimental data [10] and the TENDL-2017 data [16]. Our result is consistent with the experimental data, though the peak of the TENDL-2017 data shifts to the lower energy and underestimates above 30 MeV.

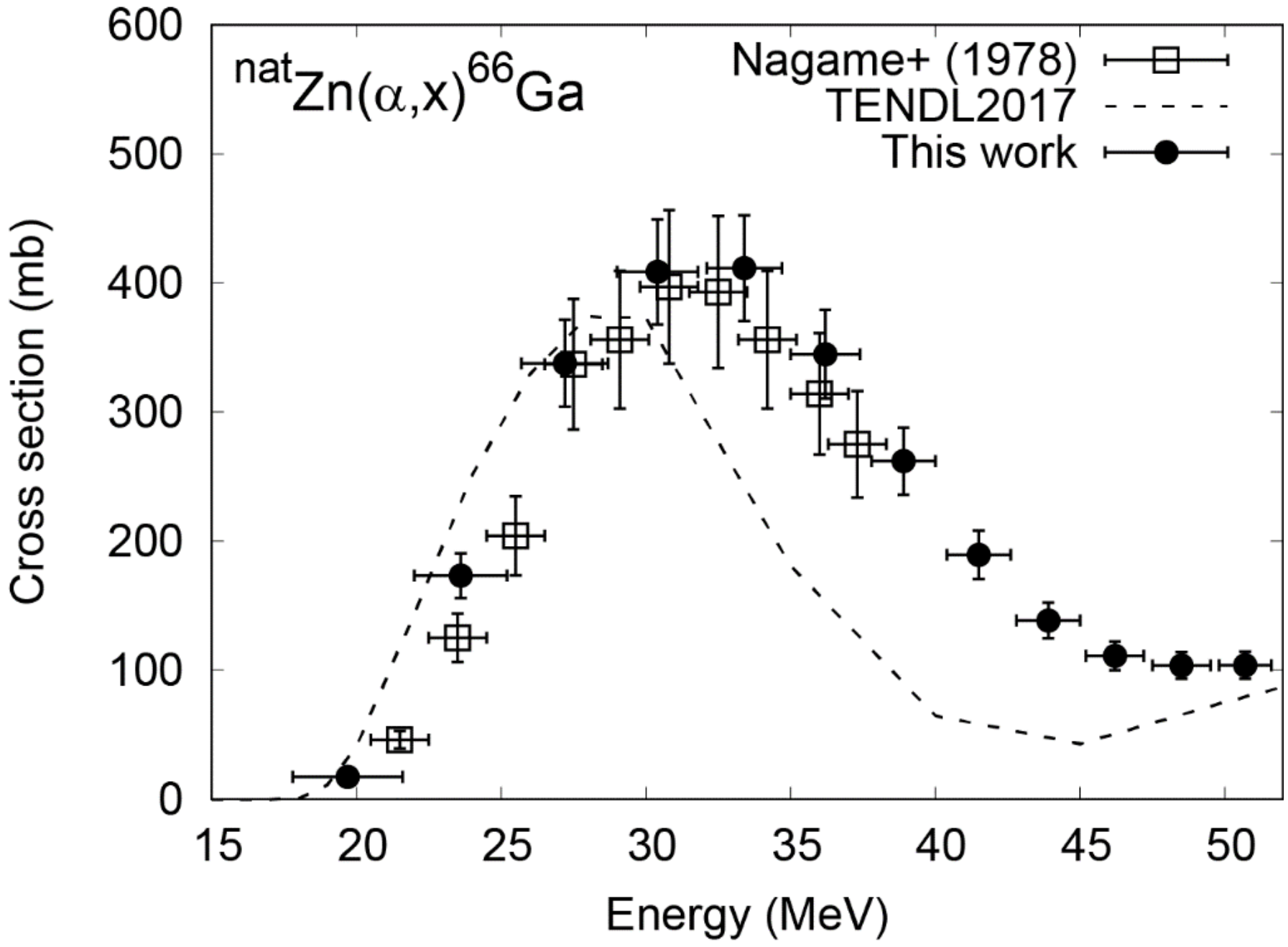

Fig. 4. Measured excitation function of the $^{nat}Zn(\alpha,x)^{66}Ga$ reaction with the previous experimental data [10] and the TENDL-2017 data [16].

### 3.4 $^{67}$Ga production

The cross sections for the $^{67}$Ga production were derived from the measurement of the 300.217-keV γ-line ($I_\gamma = 16.64\%$) from the $^{67}$Ga decay ($T_{1/2}$ = 3.2617 d). The parent nuclide $^{67}$Ge ($T_{1/2}$ = 18.9 min) is considered to decay to $^{67}$Ga in a cooling time of 22 hours. The γ-lines at 93.310 keV ($I_\gamma = 38.81\%$) and 184.576 keV ($I_\gamma = 21.410\%$) from $^{67}$Ga were not used in the present analysis to avoid possible contribution of $^{67}$Cu ($T_{1/2}$ = 61.83 h) with $I_\gamma = 16.10\%$ and 48.7%, respectively. On the contrary, the intensity of the 300.217-keV γ-line from $^{67}$Cu is relatively small ($I_\gamma = 0.797\%$) and possibly negligible. The negligible contribution could be confirmed to find a little difference among the cross sections derived from those γ-lines. The result is shown in Fig. 5 in comparison with the experimental data [9,10] and the TENDL-2017 data [16]. The peak amplitude is much larger than the other data. The peak position is nearly consistent with the experimental data, however, shifted to the higher energy side than the TENDL-2017 data. This tendency is the same as the case of the $^{nat}Zn(\alpha,x)^{66}Ga$ reaction.

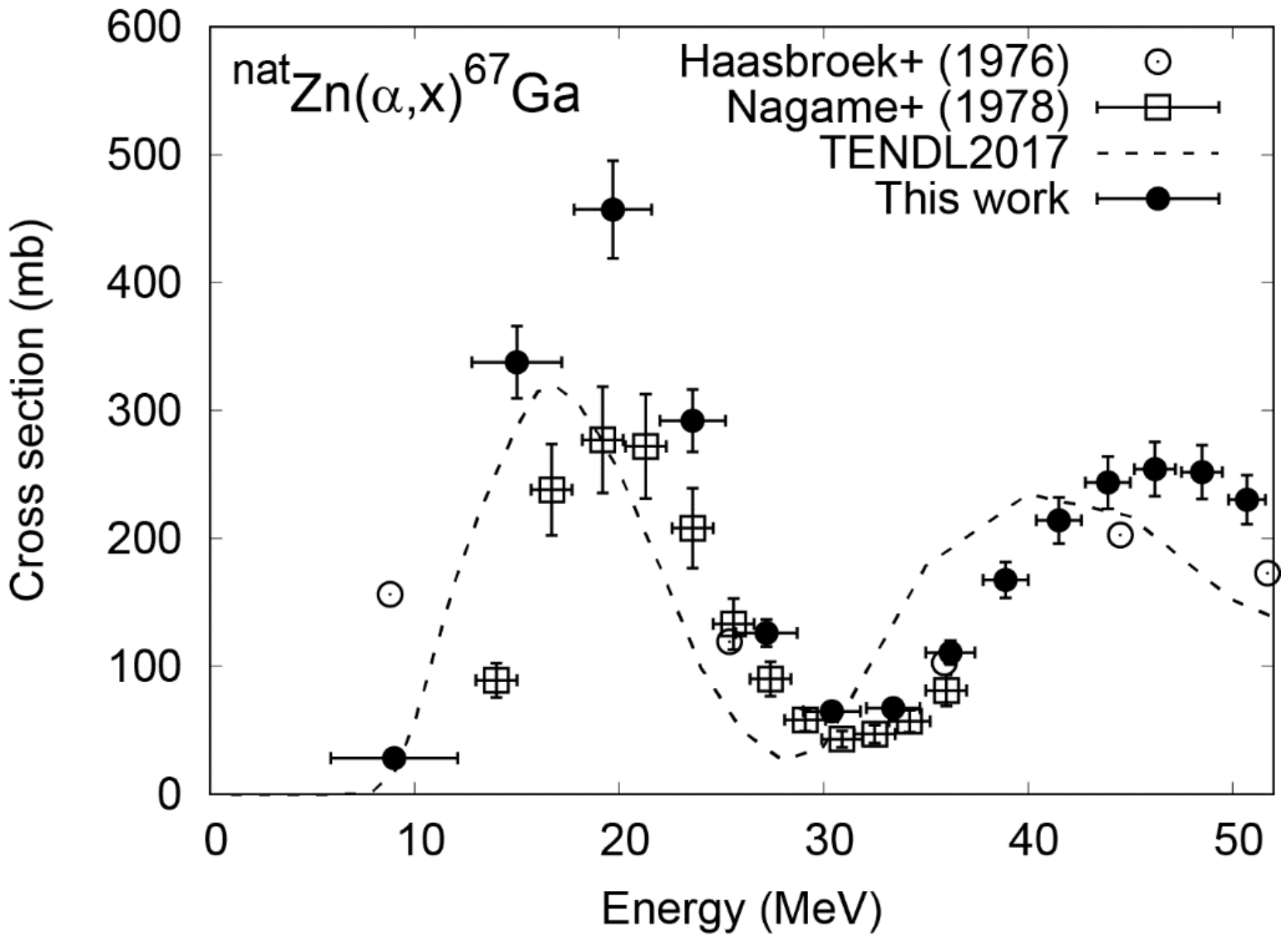

Fig. 5. Measured excitation function of the $^{nat}Zn(\alpha,x)^{67}Ga$ reaction with the previous experimental data [9,10] and the TENDL-2017 data [16].

### 3.5 Integral yield of $^{68}$Ge

The integral yield of $^{68}$Ge was estimated using Eq. (2) with the stopping powers $S(E)$ obtained by the SRIM code [12]. The cross sections of the $^{nat}$Zn(α,x)$^{68}$Ge reaction were interpolated by the spline fitting of the measured data in this work. The derived integral yield is shown in Fig. 6 together with experimental data earlier [9,10]. Our result is larger than the other data above 30 MeV as expected from the measured cross section data.

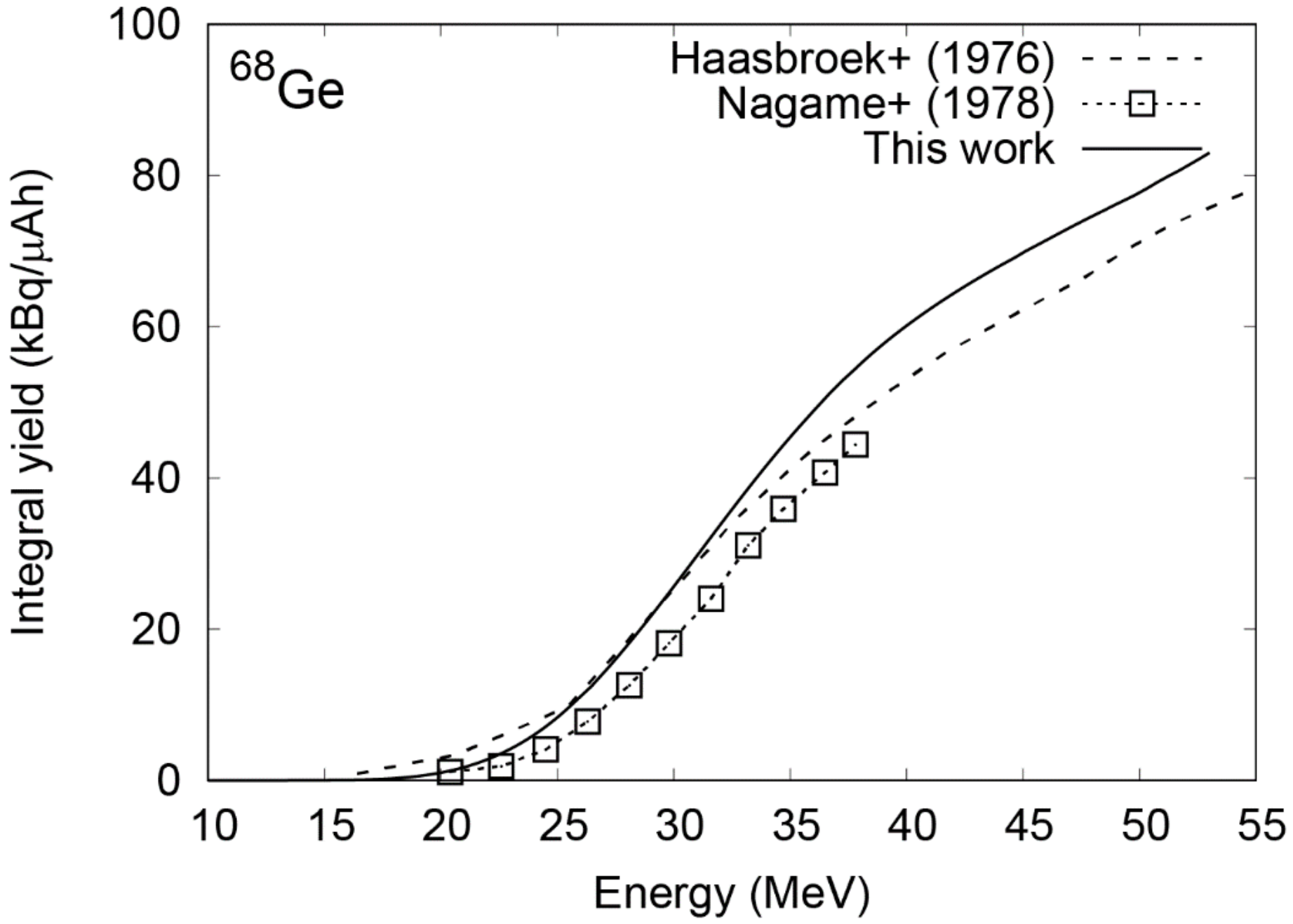

Fig. 6. Integral yield of $^{68}$Ge with the previous experimental data [9,10].

## 4. Summary

We measured the excitation function for the α-induced reactions on $^{nat}Zn$ to produce $^{68}Ge$, which is a generator of a PET radioisotope, $^{68}Ga$. The excitation function of the $^{nat}Zn(\alpha,x)^{68}Ge$ reaction was measured up to 50.7 MeV and found to be almost consistent with previous experiments. The integral yield of $^{68}Ge$ was derived from the measured cross sections and found to be larger than the previous experimental data above 30 MeV. The production cross sections of the long-lived isotopes, $^{69}Ge$ and $^{66,67}Ga$, are also measured.

## Acknowledgements

This work was performed at the RI Beam Factory operated by the RIKEN Nishina Center and CNS, University of Tokyo. This work was supported by JSPS KAKENHI Grant Number 17K07004.